# Malware Detection Module using Machine Learning Algorithms to Assist in Centralized Security in Enterprise Networks

Priyank Singhal
*Student, Computer Engineering*
*Sardar Patel Institute of Technology*
*University of Mumbai*
*Mumbai, India*

Nataasha Raul
*Research Scholar*
*Sardar Patel Institute of Technology*
*University of Mumbai*
*Mumbai, India*

**Abstract**

Malicious software is abundant in a world of innumerable computer users, who are constantly faced with these threats from various sources like the internet, local networks and portable drives. Malware is potentially low to high risk and can cause systems to function incorrectly, steal data and even crash. Malware may be executable or system library files in the form of viruses, worms, Trojans, all aimed at breaching the security of the system and compromising user privacy. Typically, anti-virus software is based on a signature definition system which keeps updating from the internet and thus keeping track of known viruses. While this may be sufficient for home-users, a security risk from a new virus could threaten an entire enterprise network.

This paper proposes a new and more sophisticated antivirus engine that can not only scan files, but also build knowledge and detect files as potential viruses. This is done by extracting system API calls made by various normal and harmful executable, and using machine learning algorithms to classify and hence, rank files on a scale of security risk. While such a system is processor heavy, it is very effective when used centrally to protect an enterprise network which maybe more prone to such threats.

## 1. Introduction

Malware, short for malicious software, consists of programming (code, scripts, and other content) designed to disrupt operation or gather information that leads to loss of privacy, gain unauthorized access to system resources, and other abusive behaviour [1]. It is a general term used to define a variety of forms of hostile, intrusive, or annoying software or program code. Any software is classified as malware based on the intent of the maker rather than any particular feature. Malware includes computer viruses, worms, Trojan horses, spyware, dishonest adware, crime-ware, most rootkits, and other malicious and unwanted software or program [2].

Symantec published a report in 2008 that "the release rate of malicious code and other unwanted programs may be exceeding that of legitimate software applications." According to F-Secure, "As much malware produced in 2007 as in the previous 20 years altogether." [3]. While these may mean nothing to the average home user, these statistics are alarming keeping in mind the financial implications of such threats to enterprises in case such threats penetrate and compromise the large volumes of data stored and transacted upon.

Since the rise of widespread Internet access, malicious software has been designed for a profit, for examples forced advertising. Since 2003, the majority of viruses and worms have been designed to take control of users' computers for black-market exploitation. Spyware are programs designed to monitor users' web browsing and steal private information. Spyware programs do not spread like viruses, but are installed by exploiting security holes or are packaged with user software [4] [5].

Clearly, there is a very urgent need to find, not just a suitable method to detect infected files, but too build a smart engine that can detect new viruses by studying the structure of system calls made by malware.

## 2. Current Antivirus Software

Antivirus software is used to prevent, detect, and remove malware, computer viruses, computer worm, Trojan horses, spyware and adware. A variety of strategies are typically employed by the anti-virus engines. Signature-based detection involves searching for known patterns of data within executable code. However, it is possible for a computer to be infected with new virus for which no signatures exist [6]. To counter such "zero-day" threats, heuristics can be used, that identify new viruses or variants of existing viruses by looking for known malicious code. Some antivirus can also make predictions by executing files in a sandbox and analysing results.

Often, antivirus software can impair a computer's performance. Any incorrect decision may lead to a security breach, since it runs at the highly trusted kernel level of the operating system. If the antivirus software employs heuristic detection, success depends on striking the correct middle point between false positives and negatives. Today, malware may no longer be executable files. Powerful macros in Microsoft Word could also present a security risk. Traditionally, antivirus software heavily relied upon signatures to identify malware. However, because of newer kinds of malware, signature-based approaches are no longer effective [7].

Although standard antivirus can effectively contain virus outbreaks, for large enterprises, any breach could be potentially fatal. Virus makes are employing "oligomorphic", "polymorphic" and, "metamorphic" viruses, which encrypt parts of themselves or modify themselves as a method of disguise, so as to not match virus signatures in the dictionary [8].

Studies in 2007 showed that the effectiveness of antivirus software had decreased drastically, particularly against unknown or zero day attacks. Detection rates have dropped from 40-50% in 2006 to 20-30% in 2007. The problem is significantly greater due to the novel methods of virus makers. Testing on major virus scanners shows that none provide 100% virus detection. The best ones provided as high as 99.6% detection, while the lowest provided only 81.8% in tests conducted in February 2010 [25]. All virus scanners produce false positive results as well, identifying benign files as malware.

## 3. Our Approach

As we have seen, current antivirus engine techniques are not optimum in detecting viruses in real time. They may be useful in controlling viruses once they infect systems, which is again, fateful for enterprises [9] [10]. This research is thus aimed at a central solution that works at the firewall level of the enterprise network. The complete system diagram is shown in Figure 1 and our process diagram is shown in Figure 2.

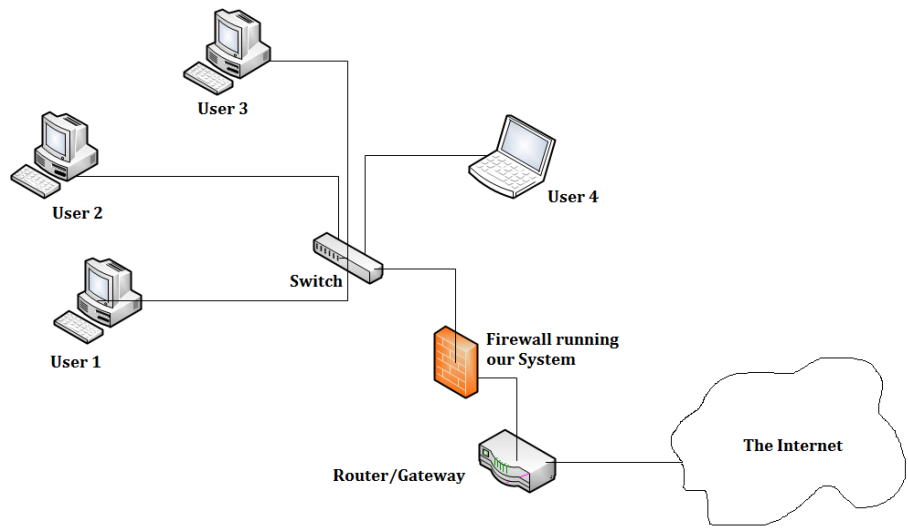

Figure 1: Network Diagram of the entire system

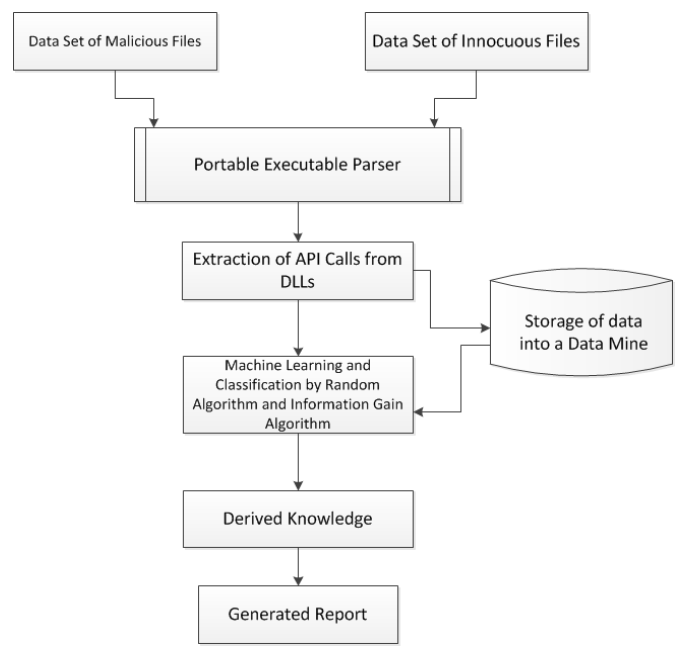

Figure 2: Process Diagram of our System

**Portable Executable (PE)**

This format is a file format for executables, object code and DLLs, used commonly in the Windows operating systems [23]. The term "portable" refers to the format's versatility in numerous environments of operating system software architecture. The PE format is essentially a data structure encapsulating necessary information for the Windows OS loader to manage the wrapped executable code. It primarily includes dynamic library references for linking, API export and import tables, resource management data, etc.

A Portable Executable file consists of a number of sections that indicate to the dynamic linker the mapping of the file in the memory. An executable image consists of several different regions, each of which requires different memory protection. The Import address table (IAT), is used to lookup tables when the application is calling a function in a different module. This is because a compiled program does not recognize the memory location of the libraries it depends and an indirect jump is required if an API call is made. The dynamic linker thus loads modules and joins, while writing physical addresses into the IAT, such that they point to the memory locations of corresponding library functions.

In our research, we extracted the PE Header from numerous infected and normal executables and using the IAT, extracted various API calls and stored them into a data mine [11] [12]. We then derived Information Gain (IG) for each function.

**Algorithm for Information Gain:**

The entropy of a variable X is defined as:

$$H(X) = -\sum_i P(X_i) \log_2(P X_i)$$

Where in H(P), the P(X) is as follows:

$$P(X_i) = \frac{Number\ of\ \cdot PE\ \cdot with\ \cdot x_i \cdot as\ \cdot certain\ API}{Total\ \cdot number\ of\ \cdot PE}$$

And the entropy of X after observing values of another variable Y is defined as:

$$H(X|Y) = -\sum_j P(Y_j) \sum_i P(X_i|Y_j) \log_2(P(X_i|Y_j))$$

The amount by which the entropy of X decreases reflects additional information about X provided by Y is called information gain, given by:

IG(X | Y) = H(X) - H(X | Y)

Machine learning, a branch of artificial intelligence, is a scientific discipline concerned with the design and development of algorithms that allow computers to evolve behaviours based on empirical data, such as from sensor data or databases [14]. A learner can take advantage of data to capture characteristics of interest of their unknown underlying probability distribution. A major focus of the research in machine learning is to learn to recognize complex patterns and to automatically make intelligent decisions based on the collected data [15].

Further, we apply the Random Forest Algorithm (RFA) [16]. This is a machine learning classification algorithm to construct the classifier to detect malware. A Random Forest is a classifier that is comprised of a collection of decision tree predictors. Each individual tree is trained on a partial, independently sampled, set of instances selected from the complete training set. The predicted output class of a classified instance is the most frequent class output of the individual trees [17] [18].

**4. Obtained Results**

To determine whether our method can provide successful results, we extracted data from over 5000 executables. These have been a combination of normal and infected files [19] [22] [24]. The first step was to create a hash map of all the executables and functions (Figure 2). After that, the information gain algorithm is used to choose only the top 80% of the functions (Figure 3), which are most likely to be present in harmful files [20]. The Information Gain is further corrected by using this formula:

$$IG(X) = IG(X) \pm \left[\frac{\sum_{i=0}^{n} IG(X_i)}{n}\right]$$

This formula helps in correcting the error by adding or subtracting the average value from the information gain value calculated. This is similar to the error correction using a standard deviation.

The results of the same are shown below:

| # | Win32.FunLove.40 | Win32.HLLP.Hanta | Worm.Win32.Loves | Joke.Win32.Zappa | Virus.Win32.Akez.e | Virus.Win32.Arcer.e | Virus.Win32.Arch.a | Virus.Win32.Aris.ex | Virus.Win32.AutoW |
|---|---|---|---|---|---|---|---|---|---|
| Is Virus? | Yes | Yes | Yes | Yes | Yes | Yes | Yes | Yes | Yes |
| ExitProcess (1) | 1 | 1 | 1 | 1 | 1 | 1 | 1 | 1 | 0 |
| GetModuleFileNa... | 1 | 1 | 0 | 1 | 0 | 1 | 1 | 1 | 0 |
| FreeEnvironment... | 1 | 0 | 0 | 0 | 0 | 0 | 0 | 0 | 0 |
| GetEnvironmentS... | 1 | 0 | 0 | 0 | 0 | 0 | 0 | 0 | 0 |
| FreeEnvironment... | 1 | 0 | 0 | 0 | 0 | 0 | 0 | 0 | 0 |
| GetEnvironmentS... | 1 | 0 | 0 | 0 | 0 | 0 | 0 | 0 | 0 |
| WideCharToMulti... | 1 | 0 | 0 | 1 | 0 | 1 | 1 | 0 | 0 |
| GetCPInfo (8) | 1 | 1 | 0 | 0 | 0 | 1 | 1 | 1 | 0 |
| GetACP (9) | 1 | 1 | 0 | 0 | 0 | 0 | 0 | 1 | 0 |
| GetOEMCP (10) | 1 | 0 | 0 | 0 | 0 | 0 | 0 | 0 | 0 |
| SetHandleCount... | 1 | 0 | 0 | 0 | 0 | 0 | 0 | 0 | 0 |
| GetFileType (12) | 1 | 0 | 0 | 1 | 0 | 1 | 1 | 0 | 0 |
| RtlUnwind (13) | 1 | 0 | 0 | 1 | 0 | 1 | 1 | 0 | 0 |
| UnhandledExcep... | 1 | 0 | 0 | 0 | 0 | 1 | 0 | 0 | 0 |
| WriteFile (15) | 1 | 1 | 0 | 1 | 0 | 1 | 1 | 1 | 1 |
| HeapFree (16) | 1 | 0 | 0 | 0 | 0 | 0 | 0 | 0 | 0 |
| HeapAlloc (17) | 1 | 0 | 0 | 0 | 0 | 0 | 0 | 0 | 0 |
| GetProcAddress (... | 1 | 1 | 1 | 1 | 0 | 1 | 1 | 1 | 0 |
| LoadLibraryA (19) | 1 | 1 | 1 | 1 | 0 | 1 | 0 | 1 | 0 |
| GetLastError (20) | 1 | 0 | 0 | 1 | 0 | 1 | 1 | 1 | 0 |

Figure 3: Hash Map of EXEs and API Functions

After running the information gain algorithm, these are the top functions:

| FunctionID | FunctionName | InfoGain |
|---|---|---|
| 1 | ExitProcess | 0.149301615453... |
| 822 | memcpy | 0.146714721368... |
| 1030 | _wcsicmp | 0.144179585504... |
| 786 | RegOpenKeyExW | 0.141365262884... |
| 891 | free | 0.123547055747... |
| 20 | GetLastError | 0.123511293485... |
| 888 | malloc | 0.120886020102... |
| 1039 | _wcsnicmp | 0.120154210826... |
| 787 | RegQueryValueE... | 0.1143423419244 |
| 526 | GetCommandLin... | 0.114333039078... |
| 565 | LoadStringW | 0.112130641786... |
| 652 | WriteConsoleW | 0.112074771986... |
| 26 | GetCommandLineA | 0.111570900397... |
| 28 | GetModuleHandl... | 0.103585190858... |
| 226 | VirtualAlloc | 0.101959007398... |
| 2053 | SetThreadUILan... | 0.100975548287... |
| 1918 | GetFileAttributesW | 0.092825662330... |
| 225 | VirtualFree | 0.092221609655... |
| 372 | SysFreeString | 0.088584587914... |
| 2008 | _wcmdln | 0.087853774626... |
| 651 | CreateFileW | 0.087731867061... |

Figure 4: Information Gain values of API Functions

Using this data, we run the Random Forest Algorithm, yielding the following functions:

| | | |
|---|---|---|
| Total Instances | 4500 | |
| Correctly Classified Instances | 4470 | 99.5556 % |
| Incorrectly Classified Instances | 30 | 0.4444 % |

Table 1: Experiment Results

| Algorithm | TP | FP | DR | ACY |
|---|---|---|---|---|
| **Decision Tree** | 0.9 | 0.1 | 90 % | 90 % |
| **Naive Bayes** | 0.95 | 0.05 | 95 % | 95% |
| **Random Forest** | 0.97 | 0.03 | 97 % | 97% |
| **Our Proposed Method** | 0.996 | 0.003 | 99% | 98% |

## 5. Conclusion

In this research, we have proposed a malware detection module based on advanced data mining and machine learning. While such a method may not be suitable for home users, being very processor heavy, this can be implemented at enterprise gateway level to act as a central antivirus engine to supplement antiviruses present on end user computers. This will not only easily detect known viruses, but act as a knowledge that will detect newer forms of harmful files. While a costly model requiring costly infrastructure, it can help in protecting invaluable enterprise data from security threat, and prevent immense financial damage.

**References**

[1] http://www.us-cert.gov/control_systems/pdf/undirected_attack0905.pdf

[2] "Defining Malware: FAQ". http://technet.microsoft.com. Retrieved 2009-09-10.

[3] F-Secure Corporation (December 4, 2007). "F-Secure Reports Amount of Malware Grew by 100% during 2007". Press release. Retrieved 2007-12-11.

[4] History of Viruses. http://csrc.nist.gov/publications/nistir/threats/subsubsection3_3_1_1.html

[5] Landesman, Mary (2009). "What is a Virus Signature?" Retrieved 2009-06-18.

[6] Christodorescu,M., Jha, S., 2003. Static analysis of executables to detect malicious patterns. In: Proceedings of the 12th USENIX Security Symposium. Washington .pp. 105-120.

[7] Filiol, E.,2005. Computer Viruses: from Theory to Applications. New York, Springer, ISBN 10: 2-287-23939-1.

[8] Filiol, E., Jacob, G., Liard, M.L., 2007: Evaluation methodology and theoretical model for antiviral behavioral detection strategies. J. Comput. 3, pp 27–37.

[9] H. Witten and E. Frank. 2005. Data mining: Practical machine learning tools with Java implementations. Morgan Kaufmann, ISBN-10: 0120884070.

[10] J. Kolter and M. Maloof, 2004. Learning to detect malicious executables in the wild. In Proceedings of KDD'04, pp 470-478.

[11] J. Wang, P. Deng, Y. Fan, L. Jaw, and Y. Liu, 2003.Virus detection using data mining techniques. In Proceedings of IEEE International Conference on Data Mining.

[12] Kephart, J., Arnold, W., 1994. Automatic extraction of computer virus signatures. In: Proceedings of 4th Virus Bulletin International Conference, pp. 178–184.

[13] L. Adleman, 1990. An abstract theory of computer viruses (invited talk). CRYPTO '88: Proceedings on Advances in Cryptology, New York, USA. Springer, pp: 354–374.

[14] Lee, T., Mody, J., 2006.Behavioral classification. In: Proceedings of European Institute for Computer

Antivirus Research (EICAR) Conference.

[15] Lo, R., Levitt, K., Olsson, R., 1995: Mcf: A malicious code filter. Comput. Secur. 14, pp.541–566.

[16] M. Schultz, E. Eskin, and E. Zadok, 2001.Data mining methods for detection of new malicious executables. In Security and Privacy Proceedings IEEE Symposium, pp 38-49.

[17] McGraw, G., Morrisett, G.,2002 : Attacking malicious code, report to the infosec research council. IEEE Software. pp. 33–41.

[18] P. Szor, 2005.The Art of Computer Virus Research and Defense. New Jersey, Addison Wesley for Symantec Press. ISBN-10: 0321304543.

[19] Rabek, J., Khazan, R., Lewandowski, S., Cunningham, R., 2003. Detection of injected, dynamically generated, and obfuscated malicious code. In: Proceedings of the 2003 ACM Workshop on Rapid Malcode, pp. 76–82.

[20] S. Hashemi,Y. Yang, D. Zabihzadeh, and M. Kangavari, 2008.Detecting intrusion transactions in databases using data item dependencies and anomaly analysis. Expert Systems, 25,5,pp 460–473. DOI: 10.1111/j.1468-0394.2008.00467.x

[21] Sung, A., Xu, J., Chavez, P., Mukkamala, S., 2004.Static analyzer of vicious executables (save). In: Proceedings of the 20th Annual Computer Security Applications Conference. IEEE Computer Society Press,ISBN 0-7695-2252-1,pp.326-334.

[22] Virus dataset, Available from: http://virussign.com/

[23] Y. Ye, D. Wang, T. Li, and D, Ye. 2008. An intelligent pe-malware detection system based on association mining. In Journal in Computer Virology, 4, 4, pp 323–334. DOI 10.1007/s11416-008-0082-4.

[24] Zakorzhevsky, 2011. Monthly Malware Statistics. Available from: http://www.securelist.com/en/analysis/204792182/Monthly_Malware_Statistics_June_2011.

[25] Dan Goodin (December 21, 2007). "Anti-virus protection gets worse". *Channel Register*. Retrieved 2011-02-24.